\documentclass[german,11pt]{article}%
\usepackage{babel}
\usepackage[latin1]{inputenc}
\usepackage{amsmath}
\usepackage{graphicx}%
\usepackage{amsfonts}%
\usepackage{amssymb}

\begin{document}

\title{Complementarity of Process and Substance}
\author{Hartmann R\"{o}mer\\Physikalisches Institut der Universit\"{a}t Freiburg\\Hermann-Herder-Str. 3\\79104 Freiburg}
\date{17.2.2006}
\maketitle

\bigskip

\bigskip

\bigskip

To my son Christoph

(1975-2004)

\bigskip

\textbf{Abstract}

Process Philosophy endeavours to replace the classical ontology of substances
by a process ontology centered on notions of changes and transitions. We
argue, that the substantial and processual approach are mutually
complementary. Here, complementarity is to be understood in the sense of a
``Generalized Quantum Theory'', which is not restricted to physical phenomena.
From this point of view, restricting oneself to either substance or process
ontology would be as ill-advised as exclusively relying on position or
momentum observables in physics. A new view on Zeno's paradox lends itself.
The meaning of an ``internal energy observable'', complementary to inner time,
and its relationship to ``akategorial states'' of the human mind will also be discussed.

\bigskip

\section{Introduction}

It is very difficult, if not impossible, to imagine a motion simultaneously
both as a unified process and as a sequence of intermediate positions.
According to a well known paradox of Zeno the Eleatic, a flying arrow seems to
freeze in its motion when attention is focussed to its momentary position at
any given instant of time, noticing that it never occupies more space than
given by its length.

The Eleate Parmenides contests the reality of all kinds of motion and change,
reducing them to the status of mere illusions. For Plato, ideas are
essentially characterized by not being subject to temporality and change, and
they are the only worthy objects of pure philosophical contemplation.

In sharp contrast hereto, motion and change are central notions in the
thinking of Aristotle. However, he remains indebted to Plato insofar as he
understands motion exclusively from its incipient and final states, whereas
the precise way in which the transition between these states is achieved is
not \textsl{described}.

On the other hand, the presocratic philosopher Heraclite from Ephesos
attributes reality only to the flow of motion, pushing states of rest down to
a kingdom of semblance and illusion.

The question of the ontological status of time and change, for which Heraclite
and Plato mark antipodal positions has ever since been a principal subject of
European philosophy. It is not possible at this place to give an oversight,
however sketchy, of the positions which are possible and have been adopted on
this matter. It is certainly fair to say, that positions closer to Parmenides
or Plato have been advocated more frequently and more influentially. The
Aristotelian point of view is just one particularly prominent example for this.

More recently, the antagonism between the Parmenidean and Heraclitean position
has again become an explicit subject of philosophical thought under the
headings \textit{''Substance Ontology''} versus\textit{ ''Process Ontology''.}
\ The so-called \textit{Process Philosophy} \cite{Prozessphilosophie}
criticizes an alleged one-sided preference for the substantial point of view
and calls for a stronger emphasis on processual elements in ontology.

Process Philosophy usually and with a good deal of justification refers to
A.N. Whitehead \cite{whitehead} as to one of their founding fathers. Indeed,
Whitehead's philosophy is centered around notions of change, becoming and evolution.

In this study, we shall try to demonstrate that the difficulty, which becomes
apparent in Zeno's paradox has its origin in a complementary relationship of
the substantial and the processual points of view.

\textit{Complementarity} as we mean it here has been introduced by Niels Bohr
as a notion of quantum physics. Quantum observables like position and momentum
(or velocity) are called \textit{complementary,} if it is impossible\textit{
}to\textit{ }ascribe sharp values of arbitrary precision simultaneously to
both of them. Already Bohr himself repeatedly pointed out that complementarity
is a very general structure in the mutual relationship of different notions or
approaches, which should apply and be relevant far beyond the realm of physics.

Elsewhere \cite{schwache QTh} we proposed a formal framework, called
''\textit{Weak Quantum Theory'', }which might also be denoted as
\textit{''Generalized Quantum Theory''}. \textsl{In spite of its name,
Generalized Quantum Theory is not a physical theory but a theory of systems in
a very general sense. It provides} \textsl{a wide and flexible framework
}within which it is possible to talk about complementarity (and entanglement)
in a well defined and not merely metaphorical way \textsl{and predict their
existence }far beyond the range of phenomena accessible to a description in
terms of physics.

Generalized Quantum Theory shares with ordinary quantum theory the fundamental
notions of ''system'', ''state'' and ''observable''. The structure of
Generalized Quantum Theory should be realized whenever observations have an
essential and inevitable influence on the state of a system. This is clearly
true in an exemplary way for the human mind as seen from the inner perspective
of self observation. \textsl{This is the application of Generalized Quantum
Theory which we primarily have in mind in this study. }We shall relate the
incompatibility of substantial and processual approaches to a complementarity
of certain mental observables of time and transition .

In more detail, we shall proceed in the following way:

In section 2 we give a simplified description of Generalized Quantum Theory in
order to provide the necessary background to follow our subsequent arguments.
The significance of Generalized Quantum Theory is illustrated by mentioning
some of its applications, which have been worked out elsewhere \cite{schwache
QTh},\cite{Zenon},\cite{timewqth}.

Section 3 elaborates on the notion of observables in Generalized Quantum
Theory \textsl{in particular pertaining to the human mind}. A comparison with
Alexius Meinong's theory of objects (''Gegenstandslehre'') may be helpful to
emancipate oneself from an unjustified narrow preconception prejudiced by the
notion of observables in physics. In the spirit of Meinong,
\textit{intentionality }and \textit{perspectivity }are emphasized as essential
features of observables, whose creation or identification will be described as
a genuine creative act of the human mind. Also essential is the notion of
\textit{partitioning, }i.e. a separation of the totality of the world into
different parts by suitable observables. The \textit{''epistemic split'' }
into observer and an observed object is the first and most fundamental example
underlying every act of cognition.

Section 4 deals with psychic time observables and their relationship to
physical time and to time complementary observables. \ This second class of
observables should be closely related to notions of process philosophy. We
also talk about a possible resolution of Zeno's paradox and give an
explanation for the finite duration of the psychological ''now''.

\textsl{Section 5} is directly devoted to the complementarity of substance and
process. Both ways of description will turn out to be indispensable. Doggedly
sticking to one way only would be as ill advised as if an physicist would
insist in using only position or momentum observables but not both.

\textsl{Section 6 is an appendix in which a short story of Jorge Luis Borges
is quoted to demonstrate that an exaggeration of the process philosophical
stand point will lead into absurdities. On the other hand, we point out again
how deeply the human mind as seen from an internal perspective differs from a
classical physical system and that in many cases notions of process philosophy
seem to provide a more adequate description.}

\section{Generalized Quantum Theory}

\textit{Generalized Quantum Theory} \cite{schwache QTh} \cite{WQTH2}
\textsl{is not a physical theory but a general theory of systems which could
also be called non commuting system theory}. It arose from ordinary quantum
theory in algebraic form by simplification and weakening of its axioms,
leaving out everything which is specific for the world of physics. The
remaining structure is still rich enough to incorporate \textsl{and predict}
phenomena like complementarity and entanglement in a much wider framework but
still in a formally well defined sense.

The basic notions of \textit{system, state} and \textit{observable} are taken
over from ordinary quantum theory.

\begin{itemize}
\item A \textit{system} $\Sigma$ is any part of reality in its most general
sense, which can, at least in principle, be separated from the rest of the
world an can be made an object of investigation. \textsl{A system in
Generalized Quantum Theory may be very different from a system in Physics, it
may for instance, consist of an individual human mind as seen from an internal
perspective of self observation. Such systems will be our main concern in the
following sections. Another example of a Generalized Quantum system would be
formed by the mental contents of a group of researchers investigating the
Elizabethan English drama.}

\item A system has the capacity to reside in several \textit{states }$z$.
Epistemically, a state describes the degree of knowledge an observer possesses
about the system. In contrast to ordinary quantum theory, it is not assumed
that the set $\mathcal{Z}$ of all states of a system can be described by the
structure of a linear Hilbert space.

\item Every observable $A$ corresponds to a feature of the system, which can
be investigated in a (more or less) meaningful way. Let $\mathcal{A}$ denote
the totality of all observables of a system. \textsl{The most important
feature of observables in Generalized Quantum Theory, shared with ordinary
quantum theory, is the fact that the application of observables will in
general change the state of a system. Indeed, }observables can be identified
with functions on the states. This means: Every observable $A$ associates to
every state $z$ another state $A(z).$ As functions on states, observables $A$
and $B$ can be concatenated by applying first $B$ then $A$ on the states $z$.
The composite observable $AB$ is then defined by $AB(z)=A(B(z)).$ Two
observables $A,B$ are called \textit{compatible} if they commute with each
other, i.e. if $AB=BA$. Otherwise, if $AB\neq BA$, they are called \textit{
incompatible} or \textit{complementary.} In ordinary physical quantum theory,
observables can also be added and multiplied with complex numbers, and to
every observable $A$ there exists a conjugate observable $A^{\ast},$\ such
that the set $\mathcal{A}$ of all observables is endowed with a rich so-called
$C^{\ast}$-\textit{structure.} In Generalized Quantum Theory only the
multiplication described above is defined, and the totality of observables
only has the much simpler structure of a so-called \textit{semigroup}.

\textsl{''Measurement'' is another fundamental notion of both Generalized an
ordinary quantum theory. Measurement means that the investigation belonging to
an observable is really performed and that a result of the investigation is
obtained. After the measurement, the state of the system will in general
differ from the state before the measurement. For the examples of the
individual mind and the group of Elizabethan researchers it is immediately
clear that measurement and application of observables will change their
states. Already in ordinary quantum theory, the process of measurement and the
obtainment of a result are not completely describable as a dynamical process
of physics, although, of cause, they are related to a physical interaction
between a measurement device and the physical system under investigation.
Rather ''measurement'' and ''reduction of state'' are primary and largely
irreducible notions of physical quantum theory. This is also true for
Generalized Quantum Theory, which in its most general form does not even
contain a notion of dynamics.}
\end{itemize}

$\bigskip$

Generalized Quantum Theory is defined by a set of axioms. \textsl{For readers
with a background in formal mathematics we here list the most important ones.
Other readers may skip them in reading.}

\begin{itemize}
\item Associated to every observable $A$ there is a set $specA$, called
\textit{spectrum} of $A$. $specA$ is just the set of all possible outcomes of
the investigation (''measurement'') pertaining to the observable $A$.\textsl{
In Generalized Quantum Theory, }$specA$\textsl{ will not necessarily be a set
of numbers, because the outcome of an observation may be of qualitative rather
than quantitative nature}.

\item \textit{Propositions }are special observables $P$, which are reproduced
under multiplication: $PP=P,$ and whose spectra $specP$ can only contain the
two elements \quotedblbase yes `` and \quotedblbase no``. They simply
correspond to yes-no questions about the system $\Sigma$. To every proposition
$P$ there is an associated negated proposition $\bar{P}$, which is compatible
with $P$ in the sense defined above. For compatible $P_{1\text{ }}$ and
$P_{2}$ there exist a\textit{ conjunction} $P_{1}\wedge P_{2}=P_{1}P_{2}$ and
a \textit{disjunction} $P_{1}\vee P_{2}=\overline{\overline{P_{1}}%
\wedge\overline{P_{2}}}$. The laws of Boolean logic are valid for compatible propositions.

\item If $z$ is a state and if for the proposition $P$ the answer ''yes'' is
obtained in the state $z,$ then $P(z)=PP(z)=P(P(z))$ is a state, for which $P$
yields the answer ''yes'' with certainty. This is a reflection of the active,
constructive character of measurement in quantum theory as both verification
and preparation.

\item The following axiom generalizes the spectral property of observables in
ordinary quantum theory and allows to reduce all observables to propositions:
To every observable $A$ and to every element $\alpha$ in $specA$ there is an
associated proposition $A_{\alpha}$, which just means that $\alpha$ is the
result of a measurement of $A.$ Then%

\[
A_{\alpha}A_{\beta}=A_{\beta}A_{\alpha}=0\ \;\text{for }\alpha\neq\beta,\quad
AA_{\alpha}=A_{\alpha}A,\quad\bigvee_{\alpha\in specA}A_{\alpha}%
={\mathchoice{\rm1\mskip-4mu l} {\rm1\mskip-4mu l} {\rm1\mskip-4.5mu
l}\mathrm{1\mskip-5mul},}%
\]
where $0$ and ${\mathchoice{\rm1\mskip-4mu l} {\rm1\mskip-4mu l} {\rm
1\mskip-4.5mu l} {\rm1\mskip-5mu l}}$ are trivial propositions which are never
or always true respectively. The observables $A$ and $B$ are compatible if and
only if $A_{\alpha}$ and $B_{\beta}$ are compatible for all $\alpha\in specA$
and $\beta\in specB$.
\end{itemize}

The concepts of \textit{complementarity }and \textit{ entanglement }are
meaningful and important in $\ $Generalized Quantum Theory as well. For
complementary observables $A$ and $B,$ the order in which they are measured is
decisive. In Generalized Quantum Theory as well as in ordinary quantum theory
it is in general not possible to find a state $z$ for which both $A$ and $B$
have a well determined value.

\textit{Entanglement}\textsl{ is a special case of complementarity. It} can
arise, if global observables pertaining to the system as a whole are
complementary to local observable pertaining to parts of the system. In an
\textit{entangled state, }for instance in a state, in which a global
observable has a well defined value, the values of local observables are in
general not determined. However, there are typical interactionless\textit{
entanglement correlations} between the results of measurements for local
observables belonging to different parts of the system.

We explicitly stress that Generalized Quantum Theory, at least in its minimal
version presented here, does not associate quantified probabilities to the
different outcomes of the measurement of an observable $A.$ This is closely
related to the absence of any Hilbert space structure for the set of
states\ $\mathcal{Z}$.

Time is not a fundamental notion of Generalized Quantum Theory,\textsl{ and
even if time observables exist, there is not necessarily any concept of
dynamics or Hamiltonean}. Planck's constant $h$, which in ordinary quantum
theory measures the degree of non commutativity, has no privileged place in
Generalized Quantum Theory.

Generalized Quantum Theory is a universal and very flexible framework theory.
It should prove its value in situations, in which, just like in ordinary
quantum theory, measurement has an inevitable influence on the state of a
system. \textsl{This study mainly deals with systems containing human minds,
which are particularly clear examples for such a situation. }

Let us mention here some \textsl{other applications, of Generalized Quantum
Theory} which have been proposed and worked out in more or less detail:

\begin{itemize}
\item Countertranference in mentally closely bound groups of persons
\cite{schwache QTh}. The frequently reported phenomenon, that members of such
a group experience mental contents or emotions which do not seem to belong to
themselves but to other members may be described as an effect of entanglement
correlation between mental observables of different group members. The
relevant global observables are related to the degree of mental concord or
other collective observables of mood and disposition.

\item H. Walach e.a \cite{Walach} propose to explain the illusive efficiency
of homeopathy, which is otherwise hard grasp, by entanglement correlations.

\item So-called synchronistic phenomena in the sense of W. Pauli and C.G. Jung
admit an interpretation as entanglement correlations. \cite{Pauli},
\cite{jungpauli}, \cite{LucRoeWal}

\item Generalized Quantum Theory as such is timeless. In the study
\cite{timewqth} \ a scenario is described how, starting from internal time as
a form of existence as a conscious being, time observables can be identified
and their relationship to physical time can be clarified. \ In section 5 we
shall be more explicit on this matter.

\item Bistable stimuli like Necker's cube can be perceived in two different
ways. Confronted to such a bistable stimulus, conception of a person will
switch back and forth more or less regularly between the two different
interpretations. H. Atmanspacher, Th. Filk and H. R\"{o}mer\ \cite{Zenon} give
a quantitative description of the switching process in terms of Generalized
Quantum Theory. In particular, a relation between three different time
constants of perception physiology is derived in agreement with experiment.

\item In sociological systems, entanglement correlations are conceivable
between attitudes and actions of different individuals \cite{Wendt}.

\item \textsl{P. beim Graben and H. Atmanspacher \cite{BgrAtm} have
demonstrated that the structure of Generalized Quantum Theory may even be
realized in stochastic dynamical systems of classical mechanics.}
\end{itemize}

\textsl{\bigskip}

\section{\bigskip Observables}

We already mentioned that observables are associated to any feature of a
system which can be investigated in whatever way. In Generalized Quantum
Theory, a system can be quite different from a system in physics and much more
multifarious. Correspondingly, also observables will be more complex and
manifold. They are the subject of this section.

We already saw that observables are reducible to propositions or, more
precisely, to questions attributed to propositions. \ As a sentence in human
language, propositions will in general contain both nouns and verbs. Already
for this reason it would be premature to identify observables with nouns or
concepts as might be suggested by the example of physical observables like
position or momentum.

We shall in particular dwell on three characteristic features of observables:

\begin{enumerate}
\item[a)] \textit{Intentionality}, i.e. directedness on something else, as
already evident from their relationship to questions

\item[b)] \textit{Perspectivity}, because questions are posed from the
perspective of those who ask them

\item[c)] \textit{Structuring activity}, because it is by the kind and horizon
of his questions that the investigator prestructures the object of his
investigation and in a way even constitutes it.
\end{enumerate}

To a)

\textsl{We are mainly interested in observables as features of the human mind
as seen from a position of self observation.} To avoid the danger of an
erroneously narrow notion of observables, a look on the theory of objects
(''\textit{Gegenstandslehre'') } \cite{Meinong} \textit{ }of the philosopher
\textit{Alexius Meinong }(1853-1920) may be helpful. For Meinong,
\textit{object (''Gegenstand'') }is everything\textit{, }which can somehow be
given to the human mind, and he endeavors to list objects as completely as
possible. As a disciple of Franz Brentano he strongly emphasizes the
intentionality of these objects of the human mind, their directedness onto
something. Parallel to the four principal kind of activity of the human mind:
imagining, thinking, feeling and desiring, he decides between four classes of
objects (''Gegenst\"{a}nde'')

\begin{itemize}
\item \textit{Objects (in the narrower sense)}: conceptions, directed onto ''things''

\item \textit{Objectiva}: directed onto judgements or propositions

\item \textit{Dignitativa}:\textit{ }directed on values like ''good'',
''true'' or ''beautiful''

\item \textit{Desiderativa}:\textit{ }desires, obligations, purposes
\end{itemize}

The first class of objects by no means comprises only conceptions of really
existing things. On the contrary, an unbiased look at the human mind reveals
that such conceptions are rather an exception. In this context Meinong talks
about a ''prejudice in favor of the real'' (''Vorurteil zu Gunsten des
Wirklichen'') prevailing in the traditional philosophy, which is primarily
devoted to cognition.

Within the four classes Meinong differentiates between \textit{simple objects
}\ and \textit{composite objects}, composed of objects of the same or
different classes. Objects of the last three classes are always composite.
Composedness cannot be continued to infinity but ends up with objects of the
first class after a finite number of steps.

\textsl{We are now able to locate the position of observables of the human
mind in Meinong's classification: In Generalized Quantum Theory they
correspond to objectiva, which in turn may be composed of objects of all four classes.}

\bigskip

To b)

Already the name of ''observables'' \textsl{tells} that they are related to an
observer, whom one should imagine to be endowed with at least some minimal
degree of consciousness. It depends on the horizon and the perspective of the
observer what he is able and willing to observe, in other words, what are the
observables of an observed system. Perspective and horizon of the observer
will change, not the least as a result of his observations. Thereby the
totality of observables assumes a genuinely dynamic character. Both in
ordinary and in Generalized Quantum Theory, systems only arise as observed
systems. However, in ref \cite{timewqth} we showed a way to conceive the whole
of the world as a limiting system of a process of repeated enlargement of
systems. In physics, this is successfully done in \textit{Quantum Cosmology}.
The ''universe'' of Generalized Quantum Theory should be much more
comprehensive, rather like C:G: Jung's \textit{unus mundus }\cite{jungpauli},
\cite{Pauli}, which is organized by archetypes and neutral with respect to a
distinction between mind and matter.

\bigskip

To c)

We already mentioned that setting up and identifying observables must be
recognized as a crucial constitutive mental act. This is true in particular
for the \textit{partitioning} of a system $\Sigma$ into subsystems $\Sigma
_{i}$, which may be performed in many different ways under various points of
view. By this act of partitioning, the subsystems are not just registered but,
together with the total system literarily constituted.

G. Mahler \cite{Mahler} repeatedly and vigorously pointed out the key
importance of the act of partitioning into subsystems.

Partitioning is done by means of \textit{partition observables }whose
different values allow to decide between different subsystems. \ In physics,
the position observable Q is the fundamental partition observable identifying
subsystems by their different positions. Indeed, it seems to be justified to
say that the realm of physics is coincident with the range of applicability of
partitioning with respect to different positions. From this perspective, the
physical world really looks like the world of \textit{res extensae.}

If, with all due care, the whole of the world, the \textit{unus mundus, }is
conceived as a system, then the first and all-decisive partition, prior to any
act of cognition, is the \textit{epistemic split }into observer and observed.
Without such a split it is impossible to talk about knowledge to be obtained
by someone about something. The precise position of this split may be movable,
for instance in the transition from the external perspective to the internal
perspective of self observation, but the split itself can never be avoided.

One definitely has to expect that different partition observables leading to
different positions of the epistemic split may be complementary to each other.
In such situations, results of cognition obtained from different cognitional
perspectives will be incompatible. This incompatibility would not be due to
the simple fact that different perspectives cannot be assumed at the same
time, rather the results obtained from one perspective would lose their
secured validity in a different perspective.

There might even exist observables of the \textit{unus mundus} in the sense
described under point b), which are complementary to every epistemic split.
They would correspond to features of the ''universe'' which are inaccessible
from any perspective opposing an observer to something he observes.

In physics, the epistemic split arises under the name of the
\textit{Heisenberg cut }between the measuring instrument and the measured
physical object. It can be investigated to some extent by physical methods in
the theory of the measurement process in quantum physics, where a composite
system consisting of a measuring devise and an object to be measured is
analyzed. The measuring instrument and the measured object are driven into an
entangled state, and the usefulness of the measuring instrument has its origin
precisely in the ensuing entanglement correlations. The transition to the
entangled state is purely deterministic, and the typical stochastic and
indeterministic features of quantum theory only emerge after applying the
Heisenberg cut, reducing the composed system to the measuring instrument and
interpreting the measured value as a statement about the measured object.

\textsl{Interestingly enough, in physics, there is a symmetry between
measurement apparatus and measured object in the following sense: Reduction of
the state of the composite system\ to the measured object results in the same
probability distribution as reduction to the measurement devise.}

One may wonder, whether such a symmetry between observer and observed holds
true also in Generalized Quantum Theory. A symmetry of this kind would secure
the adequacy of the findings obtained by the observer and would correspond to
a tight correlation between interior and exterior world. What is observed, is
mirrored in the observer, the observed mirrors the observer, and both are part
of the same universe.

\section{Substance and Process Observables}

Generalized Quantum Theory as such does not contain any reference to time.
Also C.G. Jung's \textit{unus mundus} is entirely timeless. On the other hand,
every conscious individual is intimately bound to temporality as a mode of its
existence. Employing a distinction introduced by McTaggart \cite{McTaggart},
individual subjective time is an A-Time, which is directed from past into
future and in which presence is distinguished by a particular and unmistakable
feature of ''now''. By this, subjective A-Time differs from physical B-Time
which is of poorer structure, undirected and without a distinguished ''now''.
Rather, all points of physical time are equivalent marks on a homogenous
scale. There can be no doubt, that the subjective times of different
individuals are closely correlated with the subjective times of other
individuals and with systems in the external world like planets or clocks.

H. Primas \cite{Primas} and the author \cite{timewqth} \textit{ } proposed
different but in many respects also similar scenarios, how time could emerge
from a primordially timeless \textit{unus mundus. }Here we shall briefly
sketch the\textit{ }proposal of ref. \cite{timewqth}, which contains the
following steps:

First step: after an epistemic split of the \textit{unus mundus,} subsystems
$\Sigma_{i}$ can be identified, which correspond to conscious individuals.

Second step: In these subsystems $\Sigma_{i}$, subjective time observables
$T_{i}$ can be identified, whose values are connected by strong entanglement
correlations with observables of other subsystems. (The mechanism according to
which certain observables qualify themselves as time observables is analogous
to the emergence of a time observable via the timeless Wheeler- de Witt
equation \cite{Wheeler de Witt} of quantum cosmology.) The subjective time
observables $T_{i}$ are of A-type. So, in this scenario, the origin of time is
located in the subjective A-Times of conscious individuals.

Third step: By strong entanglement correlations, the subjective A-Times
$T_{i}$ are not only related to each others$\ $but also to observables $T_{I}$
of clocklike physical systems $\Sigma_{I}$ .

Fourth step: By a long and complicated process with many intermediate stages,
time will be more and more transported into the outside world and related to
observables of physical systems chosen in such a way that entanglement
correlations become as strict as possible. \ The physical time eventually
emerging in this procedure has lost its character as an A-Time and is left as
a B-Time of simpler structure.

Irrespectively of any concrete scenario for the emergence of time, we shall
only assume in the sequel that there are subsystems $\Sigma_{i},$which can be
identified with conscious individuals and that among the \textsl{mental}
observables of $\Sigma_{i}$ there are observables $T_{i}$ of the type of an
A-Time. Although it is not absolutely necessary, one would tend to expect
$T_{i}T_{j}=T_{j}T_{i}$ for different individuals.

\textsl{Now, considering one such subsystem} $\Sigma_{i}$ we can divide all
observables, in particular those pertaining to $\Sigma_{i}$, into two
different classes:

A) \textit{Time compatible observables} $R$ with $RT_{i}=T_{i}R$. Such
observables commute with the time observable $T_{i}.$ They are either direct
functions of $T_{i}$ or have no relation to time whatsoever. In this case, a
measurement of $R$ and an attribution of a value of time will in no way
influence or disturb each other. Examples of time compatible observables are
the position observable $Q$ \textsl{or its internal representation} \textsl{in
}$\Sigma_{i}$ and observables of shape and colour, because spacial
localizations and shapes as well as colours are completely unrelated to time.

Time compatible observables describe timeless features of a system like the
sum of angles in a triangle or Platonic ideas. \ We propose to identify such
observables with observables referring to notions of an ontology of substances
as mentioned in the Introduction.

Henceforth we shall call such observables \textit{Substance Observables. }In
the sense of section 3, Substance Observables are to be associated to nominal
sentences (in question form).

B) \textit{Time complementary observables} $S$ with $S\neq T_{i}S.$
Attributing a value to $S$ and attributing a precise mark of time are
incompatible in the sense of complementarity. A typical example of a time
complementary observable in quantum physics is given by the energy observable.
A location in time and a precise value of the energy cannot be achieved
together with arbitrary precision. Quite generally, observables will be time
complementary, if they are related to processes or changes with time.

From now on, we shall call time complementary observables \textit{Process
Observables}. They correspond to notions in an ontology of processes and are
generically related to verbal sentences (in question form).

\textsl{Even if a time variable }$T_{i}$\textsl{ and the concepts of Substance
and Process Observables are defined, this does not imply the existence of
dynamics in a system of Generalized Quantum Theory. The notion of process
applies to observables and is more general than the notion of a dymamical
equation regulating the time development of states of a system. In particular,
we cannot in general expect the existence of an observable generating all time
changes of the states. }

\textsl{One should expect that Process Observables are complementary to
Substance Observables, because the notions of endurance and change should be
incompatible in a generalized quantum system. From a formal point of view,
there is of cause a possibility of observables commuting with both }$T_{i}$
\textsl{and} \textsl{all Process Observables, but such observables, if they
existed at all would have nothing to do with the distinction between substance
and process and can be disregarded for our purposes.}

\textsl{The} complementarity of Substance Observables and Process Observables
\textsl{in} \textsl{Generalized Quantum Theory leads to} a simple resolution
of Zeno's paradox: The position of a moving body at any given time is
described by a Substance Observable like the position observable $Q,$ whereas
the motion itself is described by a Process Observable. \ The complementarity
of them explains, why the quantity of motion and the intermediate position
cannot ascribed together with arbitrary precision. In the same way, in
ordinary quantum mechanics, the notion of the orbit of a moving body, which
assigns a position to every point t of time loses its precise meaning.

Quite generally, for every change in the inner or outer world, one should
expect a complementarity between Substance Observables for intermediate states
and Process Observables for describing the phenomenon of transition itself.

In quantum mechanics, the \textsl{energy observable is privileged by being
maximally complementary} to physical time. In fact, the energy observables
functions as the generator of changes with time.

At the beginning of this section and in ref \cite{timewqth} we introduced
internal subjective time observables $T_{i}$ belonging to conscious
individuals $\Sigma_{i}$ and contended that the physical time variable $T$
arises from the observables $T_{i}$ by operationalization, externalization,
purification and structural simplification. It is now natural to wonder
whether for the individual $\Sigma_{i}$ there exists a \textsl{privileged}
\textsl{Process} Observable $E_{i}$ of energy type, whose relationship to
$T_{i}$ is analogous to the relationship between the physical time $T$ and the
physical energy observable $E$. Indeed, the notion of physical energy
developed slowly from an intuitive notion of energy as the result of a long
process of of purification and idealization. This intuitive energy notion
should help to arrive at an idea about the character of such individual energy
observables $E_{i}.$ Intuitively, the notion of energy \textsl{originally
contained} an element of will and of the ability to bring about changes. The
notion of physical energy is both sharper an narrower: it is the generator of
changes in time but it is void of any element of will or desire.

In any case, it seems to be natural to assume the existence of a
\textsl{special} Process Observable $E_{i}$ of energy type for the system
$\Sigma_{i}$. Hoping not to give rise to misunderstandings, we tentatively
call this observable \textit{mental energy. }There will be a complementarity
relation $T_{i}E_{i}\neq E_{i}T_{i}.$

This complementarity of internal time and mental energy provides an easy
explanation for an important result of the physiology of perception: The
subjective ''now'' has a finite duration of the order of magnitude of 0.03
seconds. Below this threshold of temporal distance, events cannot be arranged
in their correct temporal order. Because of the complementarity of physical
time and physical energy, an ever more precise localization in time is
possible only at the expense of an ever increasing physical energy. Likewise,
localization in subjective time should be restricted by a limited supply of
mental energy.

\textit{.}

\section{Complementarity of Substance and Process}

It should have become clear by now that, in particular for proper description
of the activities of the human mind, both notions of substance and process are
indispensable. The problem is to incorporate both into a coherent model of
thought. A remarkable attempt in this direction has been made by Atmanspacher
and Fach \cite{AtFach}, who implement ideas of William James in a the
framework of a formal model. Let us quote a passage from James' ''Principles
of Psychology'' dealing with the ''stream of thoughts''.

\textit{When the rate [of change of the subjective state] is slow we are aware
of the object of our thought in a comparatively restful and stable way. When
rapid, we are aware of the passage, a relation, a transition from it or
between it and something else. \ldots\ Let us call the resting-places the
``substantive parts'' and the places of flight the ``transitive parts'' of the
stream of thoughts. It then appears that the main end of our thinking is at
all times the attainment of some other subjective part than the one from which
we have just been dislodged. And we may say that the main use of the
transitive parts is to lead from one substantive conclusion to another.}

\textit{Now it is very difficult, introspectively, to see the transitive parts
as what they really are. If they are but flights to a conclusion, stopping
them to look at them before the conclusion is reached is really annihilating
them. Whilst if we wait till the conclusion be reached, it so exceeds them in
vigour and stability that it quite eclipses and swallows them up in a its
glare. Let anyone try to cut a thought across in the middle and get a look at
its section, and he will see how difficult the introspective observation of
the transitive act is. \ldots\ The results of this introspective difficulty
are baleful. If to hold fast and observe the transitive parts of the thought's
stream be so hard, the great blunder to which all schools are liable must be
the failure to register them, and the undue emphasizing of the more
substantive parts of the stream.}

Atmanspacher and Fach model James ''stream of thoughts'' by a Dynamic System.
The mental state $z$ is a point in a manifold of high dimensionality. The
motion of the state $z$ is driven by a potential, a function $V$, which
describes mental dispositions and modes of functioning.

The dynamics of mental activity is the assumed to be governed by a Dynamic
System, i.e. a classical ordinary differential equation of the form%

\[
\frac{dz(t)}{dt}=\nabla V(z(t))+...
\]

whose solution $z(t)$, determines the mental state at time $t.$

Particular importance must be attributed to \textit{eqilibrium states} $z_{0}%
$, in which the system can reside for an arbitrarily long span of time. They
are characterized by the condition $\nabla V(z_{0})=0$. These equilibrium
states may be divided into \textit{stable }and\textit{ unstable equilibrium
states. }After a little deviation from a stable equilibrium state $z_{s}$, the
state of the system will remain in the vicinity of $z_{s}$ whereas after a
small deviation from an unstable equilibrium state $z_{i}$ the state of the
system will migrate far away from $z_{i}$. Atmanspacher and Fach propose to
identify the stable equilibrium states with the ''substantive parts'' of
William James' ''stream of thoughts'', whereas the unstable equilibrium states
are identified with the ''transitive parts''.These latter states are called
''\textit{akategorial states'' }by Atmanspacher and Fach. Generic states are
neither substantive nor akategorial but even more unstable than unstable
equilibrium states.

It is suggestive but in no way cogent to identify the ''mental states'' of
Atmanspacher and Fach with states of the brain. In this study, we focus on the
human mind as seen from an internal perspective; the relationship between
brain and mind is not topical.

As self observation will inevitably change the state of the human mind, a
generalized quantum theoretical formalisms seems to lend itself as the
appropriate framework of description.

Our approach and the approach of Atmanspacher and Fach need not be in conflict
with each other, both of them should rather be considered as attempts to cast
a view on the human mind from different perspectives.

In the spirit of Generalized Quantum Theory, it appears natural to start out
from the distinction between Substance Observables compatible with the inner
time $T_{i}$ and Process Observables complementary to $T_{i}$ \textsl{in a
system }$\Sigma_{i}$\textsl{ describing a conscious individual.}

Substance Observables correspond to contents of the conscious human mind which
are timeless in the sense that any localization in time is irrelevant for
them. Process Observables are related to contents which resist and escape from
localization in time and are incompatible with it. The importance of such
observables for the human mind is evident: already consciousness itself is
experienced rather as a stream or flow than as an ensemble of time localizable
parts. ''Mental energy'' as described above is another important example of a
Process Observable of the human mind.

\textit{Substance }and\textit{ Process States} of the human mind can be
defined in the following way: Quite generally, an \textit{eigenstate }$z$ of
an observable $A$ is defined as a state $z$, in which it is possible to
ascribe with certainty to $A$ precisely one value in the set of its possible
values given by the spectrum of $A$. (In the notation of section 2 this can be
formalized as $A_{\alpha}(z)=z$ for an $\alpha$ in $specA$.)

Now, \textit{Substance States }are simply eigenstates of Substance Observables
and \textit{Process States }are eigenstates of Process Observables.

Substance states admit an additional localization in time, but this is in
general irrelevant as, for example, in the statement: ''This is a Square and
it is twelve o'clock''. Process States resist localization in time, and if it
is attempted nevertheless, this will change the state.

The complementarity between Substance and Process Observables is crucial for
our approach to the human mind in terms of a Generalized Quantum Theory. As
described above, it explains the finite extension of the pychological ''now''
and gives a natural resolution of Zeno's paradox. For introspection,
complementarity in the sense that one imagination may annihilate another one
is an every days's experience.

Already for the reasons given by William James, it is harder to describe a
Process State than to experience it. Process States resist the prevailing
ontology of substances. Introspectively, a Process State will escape if one
tries to get hold of it by means of Substance Observables.

Atmanspacher and Fach \cite{AtFach} devote a lot of care to a more precise
description of Process States, which they call \textit{akategorial states}.

Good examples for such states are:

\begin{enumerate}
\item Memory states, in which a process in the past is memorized as a whole

\item \quotedblbase Flow`` states, in which a person is so deeply submerged in
an activity that time seems to disappear

\item Meditative states of pure consciousness. Atmanspacher and Fach
characterize them by the following properties:

\begin{itemize}
\item Pure consciousness of universal unity

\item Absence of any localizability in space and time

\item Feeling of ultimate reality

\item Feeling of unification of opposites and of peace and harmony

\item Difficulty of conceptual description
\end{itemize}
\end{enumerate}

In our generalized quantum theoretical approach, these meditative states are
plausible candidates for eigenstates of the mental energy. In ordinary quantum
physics, the eigenstates of the time complementary energy observable are the
so-called \textit{stationary states}, time independent states resisting any
temporal localization. \ As already mentioned, mental energy also contains an
element of will and desire, and it should have eigenstates, which are not only
non localizable in time, but for which also will has come to rest and no
change in time is desired.

\textsl{In any case, we believe that we have demonstrated that Generalized
Quantum Theory is a very wide and flexible scheme, which, applied to systems
like }$\Sigma_{i}$ \textsl{provides a conceptual framework, from which views
can be cast on the human mind from a new and possibly interesting perspective.
Complementarity is predicted as a general feature and expected to hold
inevitably for Substance and Process Observables both of which are
indispensable for a complete and satisfactory description of the system.
Moreover, the resolution of Zeno's paradox and the finite duration of the
subjective ''now'' are rather surprising consequences of the general scheme of
Generalized Quantum Theory . A more detailed and in some parts perhaps even
quantitative analysis of such systems would require a concrete model for the
state space }$\mathcal{Z}_{i\text{ }}$\textsl{and for the set }$\mathcal{A}%
_{i\text{ \ }}$\textsl{of observables as functions on }$\mathcal{Z}_{i\text{
}}$\textsl{. A complete description of this kind seems to be out of reach for
a long time to come. Still, we hope to come back later to a more detailed
quantum like description of human mind and consciousness.}

\section{\bigskip Appendix: Tl\"{o}n}

Process philosophy emphasizes the importance of dynamic process oriented
notions describing motions and changes. Referring to authorities like Alfred
North Whitehead \cite{whitehead} it criticizes an allegedly exaggerated weight
put on timeless notions related to enduring substances or entities in European
Philosophy, which are blamed to hamper a proper understanding \ of phenomena
of utmost importance like renewal, innovation and creativity. \ In our
terminology, process philosophy calls for a turn away from a one-sided usage
of Substance Observables and for more attention to Process observables. It may
be an exciting and instructive exercise to try to get along without Substance
Observables as far as possible and to discard notions of time and substance.

Reflections on time and the attempt to demonstrate its illusionary character
are favored subjects of the learned Argentine writer Jorge Luis Borges
(1899-1986). In his short story \textit{\quotedblbase Tl\"{o}n, Uqbar, Orbis
Tertius``} \cite{Borges2} he describes in a wonderfully subtle, free and
playful way a Utopian planet \textit{Tl\"{o}n}, whose inhabitants do not have
any notions of substances.

Living in a world which is organized in accordance with process philosophy and
does not know about underlying enduring substances, the inhabitants of
Tl\"{o}n have no substantives in their languages. As a further natural
consequence, their philosophical views are strictly idealistic, Borges says
Berkelian. (Indeed, reductionist materialism as we know it is distinguished by
a pronounced substance ontology.) Borges describes the Tl\"{o}nians, who are
obviously completely free of the ''prejudice in favour of the real'' scorned
by Meinong, in the following way:

\textit{Hume declared for all time that while Berkeley's arguments admit not
the slightest refutation, they inspire not the slightest conviction. That
pronouncement is entirely true with respect to earth, entirely false with
respect to Tl\"{o}n. The nations of that planet are, congenitally, idealistic.
Their languages and those things derived from their language -religion,
literature, metaphysics- presuppose idealism. For the people of Tl\"{o}n, the
world is not an amalgam of objects in space; it is a heterogeneous series of
independent acts -the world is successive, temporal, but not spacial.}

About the languages of Tl\"{o}n:

\textit{There are no nouns in the conjectural} Ursprache \textit{of Tl\"{o}n,
from which its ''present day'' languages and dialects derive: There are
impersonal verbs, modified by monosyllabic suffixes (or prefixes) functioning
as adverbs. For example, there is no noun that corresponds to our word
''moon'', but there is a verb which in English would be ''to moonate'' or to
''enmoon''. ''The moon rose above the river'' is }''hl\"{o}r u fang axaxaxas
ml\"{o}'', \textit{or, as Xul Solar succinctly translates }''Upwards behind
the onstreaming it moonded''

\textit{That principle applies to the languages of the southern hemisphere. In
the northern hemisphere (about whose }Ursprache \textit{Volume Eleven contains
very little information), the primary unit is not the verb but the
monosyllabic adjective. Nouns are formed by stringing together adjectives. One
does not say ''moon''; one says ''aerial-bright above dark-round'' or ''soft-
amberish-celestial'' or any other string...}

\textit{There are famous poems composed of a single enormous word, this word
is a ''poetic object'' created by the poet. The fact that no one believes in
the reality expressed by these nouns means, paradoxically, that there is no
limit in their number.}

At this place, we should annotate that in many human languages, for instance
in Japanese, a special class of verbs is reserved to what are adjectives in
European languages. From this point of view, the difference between the
languages of the northern and southern hemisphere looks less radical in
comparison to the shared absence of nouns.

A few lines later:

\textit{Space is not conceived as having duration in time. The perception of a
cloud of smoke on the horizon and then the countryside on fire and then the
half-extinguished cigarette that produced the scorched earth is considered an
example of the association of ideas. }

The lack of understanding of the Tl\"{o}nians for enduring substances, goes so
far, that things quite common for us look scandalous for them:

\textit{Of all the doctrines of Tl\"{o}n, none has caused more uproar than
materialism. Some thinkers have formulated this philosophy (generally with
less clarity than zeal) as though putting forth a paradox. In order to make
this inconceivable thesis more easily understood, an eleventh-century
heresiarch conceived the sophism of the nine copper coins, a paradox as
scandalously famous on Tl\"{o}n as the Eleatic aporiae to ourselves. There are
many versions of that ''specious argument'', with varying numbers of coins and
discoveries; the following is the most common:}

\textit{''On Tuesday, X is walking along a deserted road and loses nine copper
coins. On Thursday, Y finds four coins in the road, their luster somewhat
dimmed by Wednesday's rain. On Friday, Z discovers three coins in the road.
Friday morning X finds two coins on the veranda of his house.''}

\textit{From this story the heresiarch wished to deduce the reality -i.e., the
continuity in time- of those nine recovered coins. ''It is absurd'', he said,
''to imagine that four of the coins did not exist from Tuesday to Thursday,
three from Tuesday to Friday afternoon, two from Tuesday to Friday morning. It
is logical to think that they in fact }did \textit{exist -albeit in some
secret way that we are forbidden to understand- at every moment of those three
periods of time.''}

\textit{The language of Tl\"{o}n resisted formulating this paradox; most
people did not understand it.}

Such a scandalous paradox must be abolished :

\textit{They explained that ''equality'' is one thing and ''identity''
another, and they formulated a sort of }reductio ad absurdum -\textit{the
hypothetical case of nine men who on nine successive nights experience a sharp
pain. Would it not be absurd, they asked, to pretend that the men had suffered
one and the same pain?}

At length and in a very elucidating way, Borges dwells on the philosophical
systems of Tl\"{o}n. \ The incapability of the Tl\"{o}nians for any notions of
substances exhibits itself in the phenomenon of \textit{hr\"{o}nir}
\textit{\footnote{The word ''\textit{hr\"{o}nir''} seems to be a free
invention of Borges with an Icelandic appeal. In Icelandic dictionaries I only
found \quotedblbase\textit{hr\"{o}nn}`` with plural \quotedblbase
\textit{hrannir}``, a poetic word with the meaning ''wave''.}, }as Borges
calls it.

\textit{Century upon century of idealism could hardly have failed to influence
reality. In the most ancient regions of Tl\"{o}n one may, not infrequently,
observe the duplication of lost objects: Two persons are looking for a pencil;
the first person finds it, but says nothing; the second person finds a second
pencil, no less real but more in keeping with his expectations. These
secondary objects are called }''hr\"{o}nir'', \textit{and they are, though
awkwardly so, slightly longer. Until recently,} hr\"{o}nir \textit{were the
coincidental offspring of distraction and forgetfulness. It is hard to believe
that they have been systematically produced for only about a hundred
years.}..\textit{.A curious bit of information}: hr\"{o}nir \textit{of the
second and third remove -}hr\"{o}nir\textit{ derived from another
}hr\"{o}n\textit{ and }hr\"{o}nir\textit{ derived from the }hr\"{o}n\textit{
of a }hr\"{o}n\textit{- exaggerate the aberrations of the first; those of the
fifth remove are almost identical; those of the ninth can be confused with
those of the second; and those of the eleventh remove exhibit a purity of line
that even the originals do not exhibit. The }hr\"{o}nir\textit{ of the twelfth
remove begin to degenerate.}...

\textit{Things duplicate themselves on Tl\"{o}n; they also tend to grow vague
or ''sketchy'', and to lose detail when they begin to be forgotten. The
classic example is a doorway that continued to exist so long as a certain
beggar frequented it, but which was lost to sight when he died. Sometimes, a
few birds, a horse, have saved the ruins of an amphitheater.}

Borges' ingenious delineation of the world of \textit{Tl\"{o}n}
\textsl{suggests the following observations:}

1) An exaggeration of process ontology will result in absurd consequences.
Some people stipulate that everything that can be captured by stable concepts
were balefully rigid and did injustice to the intimately dynamic character of
the world. Proponents of such an opinion are in danger to fall victim to the
paradox of the nine copper coins.

2) On the other hand, the world of the human mind and its products and
fictions is similar to the world of \textit{Tl\"{o}n} in many respects and
really ordered rather according to a process ontology. This already becomes
evident from the refutation of the copper coin paradox employing the example
of human pain as quoted above. Even more so, on the field of fashions, trends
and countermovements \textit{hr\"{o}nir }will be quite\textit{ }common under
the disguise of repeated rediscoveries. \textsl{So, in some situations, the
process ontology of Tl\"{o}n may be superior to the substance ontology we are
used to. }

By the way, in Borges' short story, \textit{tl\"{o}n }itself is
described\textit{ }as a product of the inventive human mind, which is about to
be replaced by the yet more complex \textit{orbis tertius}

\bigskip$\mathbf{Acknowledgement}$

I thank Harald Atmanspacher, Thomas Filk and H. Primas for critical and
helpful discussions. I am very much indebted to Klaus Jacobi and Klaus R.
Kenntemich for amicable philosophical advice. Very special thanks are due to
Georg Ernst Jacoby , whose friendship proved its worth in continuous exchange
of ideas, advice and encouragement in difficult times. I should like to
express my gratitude to my family and to all those who supported me in times
of grief.

\end{document}